\documentclass[a4paper,11pt]{article}
\pdfoutput=1 

\usepackage{jheppub} 

\usepackage[T1]{fontenc} 

\title{\boldmath Scale-Dependent Power Spectrum from Initial Excited-de Sitter Modes}

\author[a]{E. Yusofi}
\author[b,1]{M. Mohsenzadeh,\note{Corresponding author.}}

\affiliation[a]{Department of Physics, Science and Research Branch, Islamic Azad University, Tehran, Iran}
\affiliation[b]{Department of Physics, Qom Branch, Islamic Azad University, Qom, Iran}

\emailAdd{e.yusofi@iauamol.ac.ir}
\emailAdd{mohsenzadeh@qom-iau.ac.ir}

\abstract{In this paper, we calculate corrections of scalar perturbations spectra resulting from excited-de Sitter modes as the nontrivial initial states. To obtain these modes, we consider the asymptotic expansion of the Hankel functions up to the higher order of $\frac{1}{k\tau}$. Actually the Planck and WMAP data impose some constrains on the Hankel function index. These observational constraints and back-reaction effects stimulate us to use excited-de Sitter modes. Finally, we nominate these nontrivial general solutions as the fundamental mode functions during inflation and we calculate the corrected form of scale-dependent power spectrum with trans-Planckian corrections, and in de Sitter space-time limit the results reduce to the scale-invariant power spectrum.}

\begin{document}
\maketitle
\flushbottom

\section{Introduction}
\label{sec:intro}

Inflation as the most exciting scenario for cosmologists, naturally solves several puzzles of the standard big bang model \cite{inf1, inf2, inf3}. The rapid expansion of the universe during inflation is considered to be responsible for homogeneity and isotropy on the large scale structure (LSS) of the universe. On the other hand, the quantum fluctuations of the scalar (inflaton) field are generated from vacuum fluctuations during inflation, and inflation would lead to the growth of the modes of fluctuations in the accelerating phase. By this mechanism, the quantum fluctuations located outside the horizon, freezed and became the classical fluctuations. After these process, the temperature fluctuations $\frac{\Delta T}{T}$ indeed becomes the observable in the cosmic microwave background (CMB) by Sachs-Wolfe effect and the frozen fluctuations outside the horizon appear in the form of clusters of galaxies in the LSS of the universe. Almost scale-invariant power spectrum of primordial fluctuations is the most important prediction of inflation scenario \cite{per4, per5, per6, per7, per8}.\\
The amazing development of observational cosmology opens new avenues to the very early universe. The scalar spectral index $n_{s}$, tensor-to-scalar ratio $r$ \cite{ret9, ret10}, the gravitational wave, non-Gaussianity parameter $f_{NL}$ and the isocurvature perturbation in the CMB \cite{obs11} are among the most important observational probes. These probes will constrain the initial conditions of inflation model, trans-Planckian physics and alternative models to inflation, etc. One of the most important observations of the recent CMB data from Planck combined with the large angle polarization data from the WMAP, gives a strong constraint on scalar spectral index: $n_s = {0.9603} \pm {0.0073}$ at 95\% CL \cite{obs12}. Power law inflation (PLI) with $a(t)\propto t^p$ and $p > 1$ arises in the context of a canonical scalar field with an exponential potential \cite{pli13, pli14, pli15, pli16}. In \cite{pli17} a new model of power law inflation has been proposed in which the scalar spectral index, the tensor-to-scalar ratio and the non-Gaussianity parameter are in excellent agreement with Planck results. However, results of our study confirm a slow-roll PLI form of expansion for the early universe with excited-dS initial states. Non-trivial initial states for perturbations naturally accommodates the running of the tilt needed to explain the BICEP2 result \cite{Ash18}. Additionally, in \cite{map19}, by the combinations of BICEP2+WMAP 9-year data, the pLI model and the inflation model with inverse power-law potential can fit the data nicely. These interesting data can be stimulated us to slow-roll PLI for excited-dS space-time\footnote{For pure de Sitter (dS) space-time we have exponentially inflation , for excited-dS space-time we have PLI and for almost dS space-time we have slow-roll PLI. }. \\
On the other hand, recent observational data in the CMB may come from various sources during the cosmic evolution. It is believed that if all the process and transformations during cosmic expansions are linear, thus the temperature fluctuations are Gaussian. Meanwhile, any deviation from linearity in the expansion process or in the transformations between various stages of expansion process,  will influence the final observable data \cite{alf18}. As the oldest non-linear factors influencing the CMB radiation is the nontrivial initial states, that could be an important source to generate scale-dependent power spectrum and non-zero bispectrum. Moreover, because we do not know anything about the physics before inflation, a priori any excited state is as good an initial state as the vacuum state. In this paper, we shall propose the excited-de Sitter(ED) modes instead to the Bunch-Davies (BD) mode \cite{Bun19} as the nontrivial initial states \cite{man20}. We will show that this deviation from BD mode can modify power spectrum and in particular can produce scale-dependent power spectrum fit with recent observations.\\
Non-trivial initial states have been studied by many authors, including $\alpha$-vacuum \cite{alf18}, general multi-mode squeezed states \cite{non21, non22, non23, non24, non25, non26}, particle number eigen-states \cite{non27}, Gaussian and non-Gaussian initial states \cite{non28}, coherent states and $\alpha$-states \cite{non29, non30}, thermal states \cite{non31}, homogeneous initial states \cite{non32} and \emph{calm} states \cite{non33}. \\
It is also well known that the inflation can be described in approximate dS space-time \cite{per7, man34}, thus the study of dS and quantum theory of fields in this background is well motivated. For example, exploring the possible relations between the dS symmetries and the bispectrum of the fluctuations are well studied in \cite{man35} in order to set constraints on the initial fluctuations due to the dS symmetry, especially, how scale transformations and special conformal symmetries constrain the correlation functions. The scalar field in dS background is important because most of inflationary models are theorized using the scalar field \cite{man36}. In this paper we will study scalar field in the dS background with ED modes.\\
Another motivation for considering ED modes as the initial states, arise from the possibility of trans-Planckian effects \cite{tra40, tra41, tra42, tra43, tra44, tra45, tra46, tra47, tra48, tra49}. As it is known, we observe the modes that have physical momenta $k_{phys} = k/a$ which exceed the Planck energy scale $M_{Pl}$ at sufficiently far past times. The effects of trans-Planckian corrections to the modes and dynamics of it, may actually be followed by using the modified initial states. We expect our nontrivial initial states, i.e. ED modes give the trans-Planckian corrections of power spectrum up to higher order of $ H/M_{Pl}$, where $H$ is the Hubble parameter during inflation.\\
The rest of this paper proceed as follows: In Sec. 2, we first review quantum fluctuations of scalar field during inflation. In this section, we remind equation of motion, field quantization, boundary conditions for vacuum mode selection. In Sec. 3, we propose ED modes instead BD mode and to achieve this goal we consider asymptotic expansion of the Hankel functions, by generalizing Mukhanov equation and mode functions. In Sec. 4, we provide both observational and theoretical reasons for considering ED modes. Finally, we calculate scale-dependent power spectrum with using this ED modes in terms of index of Hankel function and conformal time $\tau$. Conclusions will be discussed in Sec. 6.

\section{Review of Quantum Fluctuations during Inflation}
\subsection{Equation of Motion for Scalar Field}
The following metric is used to describe the very early universe during inflation:
\begin{equation}
ds^{2}=dt^{2}-a(t)^2{d\textbf{x}}^{2}=a(\tau)^2({d\tau}^2-{d\textbf{x}}^2),
\end{equation}
where for the conformal time $\tau$ the scale factor is defined by $a(\tau)$. There are some models of inflation but the single-field inflation in which a minimally coupled scalar field (inflaton) in dS background, is usually studied in the literatures. The action for single-field models is given by:
\begin{equation}\label{act2}
S=\frac{1}{2}\int d^4x\sqrt{-g}\Big[{\cal R}-(\nabla \phi)^2-2V(\phi)\Big],
\end{equation}
where $8\pi G=\hbar=1$. We use gauge-invariant variables to avoid fictitious gauge modes by introducing comoving curvature perturbation \cite{per7},
\begin{equation}\label{gau3}
{\cal R}=\Psi+{\frac{H}{\dot{\bar{\phi}}}{\delta\phi}},
\end{equation}
where $\Psi$ is the
spatial curvature perturbation. For this gauge-invariant variable one may expand the action (\ref{act2}) to second order,
\begin{equation}\label{act4}
S=\frac{1}{2}\int d^4x a^3\frac{\dot{\bar{\phi}}^2}{H^2}\Big[\dot{{\cal R}}^2-a^{-2}(\partial_i{\cal R})^2\Big],
\end{equation}
by defining the Mukhanov variable $u=z{\cal R}$ with $z^2=a^2\frac{\dot{\bar{\phi}}^2}{H^2}$, and transitioning to conformal time $ \tau$ leads to the Mukhanov action,
 \begin{equation}\label{act5}
S=\frac{1}{2}\int d^3xd{\tau} \Big[(u')^2+(\partial_i{u})^2-\frac{z''}{z}u^{2}\Big],
\end{equation}
where the prime is the derivative with respect to conformal time. Also, we define the fourier expansion of the field $u$:
\begin{equation}
u(\textbf{x},\tau)=\frac{d^3k}{(2\pi)^{3}}\int_k u_k(\tau)e^{i\textbf{k}\cdot\textbf{x}}.
\end{equation}
 The action (\ref{act5}) leads to the Mukhanov equation (equation of motion for the mode functions $u_{k}$) satisfies \cite{per7, non29, non32},
\begin{equation} \label{muk7}  u''_{k}+(k^{2}-\frac{ z''}{z})u_{k}=0. \end{equation}

\subsection{Field Quantization and Boundary Conditions for vacuum Selection}

For quantization of the field, we promote $u$ to quantum operator $\hat{u}$,
 \begin{equation}
 \label{mod8}
u\rightarrow \hat{u}=\int\frac{d\textbf{k}^3}{(2\pi)^3}\Big( \hat{a}_\textbf{k}u_k(\tau)e^{i\textbf{k}\cdot\textbf{x}}+ \hat{a}_\textbf{k}^\dagger{u}
^{*}_k(\tau)e^{-i\textbf{k}\cdot\textbf{x}}\Big),
 \end{equation}
and the Fourier components $u_k $ are expressed via the following decomposition,
 \begin{equation}
 \label{mod9}
u_\textbf{k}\rightarrow \hat{u}_\textbf{k}=\frac{1}{\sqrt{2}}(\hat{a}_\textbf{k}u_k(\tau)+\hat{a}_{-\textbf{k}}^\dagger{u}^{*}_{-k}(\tau)).
 \end{equation}
  The canonical commutation relations for creation and annihilation operators given by:
   \begin{equation}
 \label{com10}
[\hat{a}_\textbf{k},\hat{a}_{\textbf{k}'}^\dagger]=(2\pi)^3 \delta^{3}(\textbf{k}-\textbf{k}') \\    ,\quad  [\hat{a}_{\textbf{k}}^\dagger,\hat{a}_{\textbf{k}'}^\dagger]=[\hat{a}_{\textbf{k}},\hat{a}_{\textbf{k}'}]=0.
 \end{equation}
 The \emph{normalization condition} for the quantum mode functions is can establish if and only if we have,
   \begin{equation}
  \label{nor11}
u^{*}_{k}u^{'}_{k}-u^{*'}_{k}u_{k}=-i.
 \end{equation}
 Equation (\ref{nor11}) provides one of the boundary conditions on the solutions of (\ref{muk7}) but this doesn't fix the mode functions completely. The second boundary condition comes from \emph{vacuum selection}.
 Real vacuum state for any fluctuations given by \cite{per7, non29},
   \begin{equation}  \label{vac12}
   {\hat{a}_{\textbf{k}}{\big|0\big\rangle}}=0.
 \end{equation}
First, we consider the standard choice, i.e. the Minkowski vacuum in the very early universe, $\tau\rightarrow-\infty$ or $ |k\tau|\gg1$. In this limit, we define the vacuum state by matching solutions to the Minkowski vacuum mode in the ultraviolet limit, i.e. on small scales when the mode is deep inside the horizon. This constrain fixes the mode functions completely. In this limit, we have $\frac{{z''}}{z}\rightarrow0 $  and equation (\ref{muk7}) change to special form as,
\begin{equation} \label{min13}  u''_{k}+k^{2}u_{k}=0. \end{equation}
For this case, we have a unique solution as,
\begin{equation} \label{min14}  u_{k}=\frac{e^{-ik\tau}}{\sqrt{k}}. \end{equation}
After imposing these two boundary conditions (\ref{nor11}) and (\ref{min14}), the suitable mode completely qualified on the all scales \cite{per7}.

\section{Excited-de Sitter Modes Instead Bunch-Davies Mode}
In this section, by the asymptotic expansion of the Hankel function up to the higher order of $1/k\tau$, we obtain the general solutions of  Mukhanov equation. We nominate this nontrivial solution as the fundamental mode functions during inflation. Then in section 5, we calculate power spectrum with these nontrivial modes and obtain modified form of scale-dependent power spectrum.\\
 The general solutions of mode equation (\ref{muk7}) can be written as \cite{inf2, inf3, per7, per8}: \begin{equation}
\label{Han22}
u_{k}=\frac{\sqrt{\pi \tau}}{2}\Big(A_{k}H_{\nu}^{(1)}(|k\tau|)+B_{k}H_{\nu}^{(2)}(|k\tau|)\Big), \end{equation} where
$ H_{\nu}^{(1, 2)} $ are the Hankel functions of the first and second kind, respectively \cite{inf2, per8}. But since the function  $z$ in mode equation (\ref{muk7}) is a time-dependent parameter and depends on the dynamics of the background space-time, thus finding the exact solutions of the equation (\ref{muk7})is difficult and the numerical and approximate methods are actually needed.

\subsection{Asymptotic Expansion of the Hankel Functions}
 In the far past $|k\tau|\gg1$ in the very early universe, we are authorized to use asymptotic expansions of the Hankel function up to the higher order of $\frac{1}{|k\tau|}$ as follows \cite{per7, tra50, man51},

  \begin{equation}
 \label{asm23} H_{\nu}^{(1, 2)}(|k\tau|)\rightarrow\sqrt{\frac{2}{\pi{|k\tau|}}}\big[1\pm{i}\frac{4\nu^2-1}{8|k\tau|}-\frac{(4\nu^2-1)(4\nu^2-9)}{2!(8|k\tau|)^2}\pm...\big]\times{exp}[\pm{i}(|k\tau|
 -(\nu+\frac{1}{2}))\frac{\pi}{2}],
\end{equation}
Note that, this asymptotic solution, only for $\nu=\frac{3}{2}$ reduce to the exact dS mode, which consists the first (linear) order of $\frac{1}{|k\tau|}$ and for another value of $\nu$, the modes can contain other non-linear terms of order $\frac{1}{|k\tau|}$.\\
\subsection{Mukhanov Equation and Excited-de Sitter Modes}
For the dynamical inflationary background, we have $ \frac{{z''}}{z}\neq0 $ in equation (\ref{muk7}) and it is a time-dependent value in terms of conformal time $\tau$ as following \cite{asl55},
\begin{equation} \label{zed24}
\frac{{z''}}{z}=\frac{2\alpha}{\tau^2}.
 \end{equation}
Also in addition to the variable $\tau $, the value of $ \frac{{z''}}{z}$ is depended to the Hankel function index $\nu$. Therefore, the equation (\ref{muk7}) turns to the general form
\begin{equation} \label{Muk25}  u''_{k}+(k^{2}-\frac{2\alpha}{\tau^2})u_{k}=0. \end{equation}
Variable $\alpha$ in (\ref{zed24} and \ref{Muk25}) is given by \cite{per7, asl55},
 \begin{equation} \label{alf26}  \alpha=\alpha(\nu)=\frac{4\nu^2-{1}}{8}.\end{equation}
Consequently, according to the general equation of motion (\ref{Muk25}) and the asymptotic expansion (\ref{asm23}), the general form of mode functions becomes,
 \begin{equation} \label{gen27}  u^{gen}_{k}(\tau, \nu)=A_{k}\frac{e^{-{i}k\tau}}{\sqrt{k}}\big(1-i\frac{\alpha}{k\tau}-\frac{\beta}{k^2\tau^2}-...\big)+
 B_{k}\frac{e^{{i}k\tau}}{\sqrt{k}}\big(1+i\frac{\alpha}{k\tau}-\frac{\beta}{k^2\tau^2}+...\big)
, \end{equation}
where $\beta={\alpha(\alpha-1)}/{2}$. Note that, we consider $|\tau|=-\tau$ for far past. Also, the general mode (\ref{gen27}) is a function of both variables $\tau$ and $\nu$. After making use of two boundary conditions (\ref{nor11}),(\ref{min14}) and mode function (\ref{gen27}) up to order of $\frac{1}{k^2\tau^2}$, the positive frequency solutions of the mode equation (\ref{Muk25}) becomes,
\begin{equation} \label{mod28}  u^{ED}_{k}=\frac{e^{-{i}k\tau}}{\sqrt{k}}\left(1-i\frac{\alpha}{k\tau}-\frac{\beta}{k^2\tau^2}-...\right)
. \end{equation}
We call this solutions as \emph{excited-de Sitter(ED) modes}. In addition to the above motivations, the following items can be considered as the related issues with ED modes:
\begin{itemize}
  \item Corrections obtained from previous conventional methods for power spectrum is typically of the order of $1, 2$ [40- 49]. So it is useful for us to extend the modes up to non-linear order of its parameters. This non-linearity of our ED modes appears in the both variables of it, i.e. variable $\nu$ and conformal time variable $\tau$ (In this paper we discuss this topic).
  \item Primordial Non-Gaussianity in the CMB may be come from various non-linear sources during the cosmic evolution. Any non-linear effect in the expansion process or in the initial conditions (vacuum states) may leave non-linear traces in the observable parameters of CMB \cite{alf18, Gau52, Gau53, Gau54}. So, non-linear terms of ED modes can be considered as a non-linear effect in the initial states (This issue will studied in future works).
  \item In the context of general initial states and non-Bunch-Davies mode considering the \emph{backreaction} and the \emph{tadpole} are important issues. We demand of the backreaction, constraint for energy density and the vanished tadpole of fluctuations about the correct zero mode trajectory [30, 32, 34]. So, some related issues such as the backreaction effects associated with our excited-dS modes should be discussed in detail ( In this paper we bring short discussion of this issue and we will discuss in detail in preparation work \cite{man66}).
  \item Also, compared with previous BD mode, these ED modes could be more complete solution of the general wave equations (\ref{Muk25}) for the general curved space-time(specially excited-dS space-time ), whereas BD mode is a specific solution for a specific curved space-time (i.e. dS space-time ).
\end{itemize}
\subsection{Some of Exact and Approximate Solutions}
The ED modes (\ref{mod28}) are leading to the exact solutions for the $\nu=\pm\frac{1}{2}, \pm\frac{3}{2}, \pm\frac{5}{2}, ... $ and are leading to the approximate solutions for another values of $\nu$. For the our cosmological proposes, we investigate $\nu=\frac{1}{2}, \frac{3}{2}$, $\nu\simeq\frac{3}{2}+\epsilon$ with $\epsilon<1$ and $\nu=\frac{5}{2}$.\\
\begin{itemize}
  \item For the $\nu=\frac{1}{2}$, we have from equation (\ref{zed24}), $ \frac{z''}{z}=0$. So, the mode function (\ref{mod28}) leads to exact Minkowski mode (\ref{min14}).
  \item In the case of $\nu=\frac{3}{2}$, we obtain from equation (\ref{zed24}), $ \frac{z''}{z}=\frac{2}{\tau^2}$, and the general form of the mode functions (\ref{mod28}) leads to the exact BD mode:
  \begin{equation}
 \label{Bun29} u_{k}^{BD}=\frac{1}{\sqrt{2k}}(1-\frac{i}{k\tau})e^{-ik\tau}.
\end{equation}
For this special case, we have exponentially inflation, $a(t)=e^{Ht}$ or $ a(\tau)=-\frac{1}{{H}\tau}$ with $H=constant$ for very early universe.
  \item Also, considering $\nu=\frac{5}{2}$ , we obtain from equation (\ref{zed24}), $ \frac{z''}{z}=\frac{6}{\tau^2}$. So, we have an exact excited-dS solution as,
  \begin{equation}
 \label{excited mode} u_{k}^{ED1}=\frac{1}{\sqrt{k}}(1-\frac{3i}{k\tau}-\frac{3}{k^2\tau^2})e^{-ik\tau},
\end{equation}
note that this last mode is the exact solution of Mukhanov equation (\ref{Muk25}) and it is a non-linear solution of order $\frac{1}{k^2\tau^2}$.
 \end{itemize}
Inflation starts in approximate dS space-time and basically in the very early universe with varying $H$, finding a proper mode is difficult. So, we offer the general form of excited-dS solution (\ref{mod28}) as the fundamental modes during inflation that asymptotically approaches to flat background in $\tau\rightarrow{-\infty}$ . For these fundamental modes, we will show in the next section, the best values of $\nu$ which are confirmed with the latest observational data, is $\nu=\frac{3}{2}+\epsilon$, where $0< \epsilon <1$. For this range of $\nu$, we have power law inflation (PLI) $a(t)\propto{t^p}$ with $p>1$ or $a(\tau)\propto |\tau|^{\frac{1}{2}-{\nu}}$ with ${\nu}>{3/2}$. But for $\epsilon \ll1$ we have slow-roll PLI \cite{pli17}.\\
It is important to note that if we consider mode functions depending on the Hankel function index $\nu$, BD mode is exclusively suitable only for exponentially inflation in pure de Sitter phase with $\nu={3/2}$. But for slow-roll PLI in the excited-dS space-time, it is better and more logical that we make use general ED modes (\ref{mod28}) instead BD mode (\ref{Bun29}).
\section{Some Motivations for Excited-de Sitter Modes}
\subsection{Constraint on Spectral Index from Planck and WMAP}

In this section we present Planck(2013) results for spectral index of inflation \cite{obs12} as observational motivation for ED modes.
The solution of equation (\ref{zed24}) for this case ( i.e. $\nu> 3/2$ ) leads to PLI, corresponds to
\begin{equation} \label{pli51}
 a(t)\propto t^{p} \quad or  \quad a(\tau)\propto |\tau|^{\frac{1}{2}-{\nu}}.
\end{equation}
For this case, the Hubble parameter is time-dependent variable and easy to see that from (\ref{pli51}), we obtain
\begin{equation}
 H(t)=\frac{p}{t}.
\end{equation}
In terms of variable $\nu$, we obtain $p$ as follows,
\begin{equation}
 p=\frac{(1/2-\nu)}{(3/2-\nu)},
\end{equation}
and the slow-roll parameter for PLI given by
\begin{equation}
 \epsilon=\frac{1}{p}=\frac{(3/2-\nu)}{(1/2-\nu)}.
\end{equation}
Slow-roll PLI corresponds to $\epsilon\ll1$ and $\nu=3/2 +{\epsilon}/{(1-\epsilon)}\simeq 3/2 +{\epsilon}$ \cite{asl55}. For this limit one finds,
\begin{equation}
 n_{s}-1\simeq\frac{-2}{p}=\frac{2(3-2\nu)}{2\nu-1}.
\end{equation}
Note that if we consider PLI without slow-roll assumption, one gets
\begin{equation}
 n_{s}-1=\frac{2}{1-p}= 3-2\nu.
\end{equation}
On the other hand, Planck results in combination with the large angle polarization data from WMAP requires the value of the
scalar spectral index $n_{s}$ to lie in the range $0.945\leq n_{s}\leq 0.98$ [12]. This restricts $p$ in the PLI to the range,
\begin{equation}
 38\lesssim p \lesssim 101,
\end{equation}
and equivalently we obtain for $\nu$ and $\epsilon$,
 \begin{equation}
 1.51\leq \nu \leq 1.53,
\end{equation}
\begin{equation}
 0.01\lesssim \epsilon \lesssim 0.03.
\end{equation}
Since the BD mode is used just for pure dS space-time, this above range of $\nu$ based on the most recent observations, motivate us to deviations from BD mode. Consequently, the ED modes (\ref{mod28}) proposed in this paper could be a good candidate for modifying of the standard BD mode.

\subsection{Constraint on Back-reaction}
An excited initial state can in principle affect the background inflationary dynamics (through the back-reaction of the perturbations) as well as the CMB temperature anisotropy. To this end we recall some results of cosmological perturbation theory associated with our ED modes, whose details will be presented in \cite{man66}.\\
Since, we have considered the higher non-linear order terms of $1/{k\tau}$ for ED modes to obtain corrected spectra, therefore needed that the study of back-reaction effects appear in the high-order terms associated with these excited modes. As a result, it is expected that the back-reaction effects are negligible during the near exponential expansion for BD mode i.e. $\nu=3/2$, and these effects might be important for the subsequent cosmic inflation, specially for the PLI with ${\nu}>{3/2}$. However, vacuum energy density accumulated during PLI can be important at later stages, since it increases like $1/k^{2}\tau^{2}$, which is bigger than the $1/{k\tau}$ at late time, $ |k\tau|\ll{1}$.\\ The expectation value of the EMT for inflaton field $\phi$ given by
\begin{equation}
 \label{vac4}
 T_{\gamma\delta}=\phi_{;\gamma}\phi_{;\delta}-{\frac{1}{2}}{g_{\gamma\delta}}{g^{\rho\sigma}}\phi_{;\rho}\phi_{;\sigma}.
\end{equation}
For the general ED modes (\ref{gen27}), with definition $\phi_{k}=\frac{u_{k}}{a(\tau)}$, the $T^{0}_{0}$ component (the energy or pressure density) carried by the perturbations is \cite{tan1},
$$\delta\rho_{ED}\sim\frac{1}{2a_(\tau)^{2}}\int\frac{d^{3}k}{(2\pi)^{3}}$$
\begin{equation}
 \label{vac5}
\times\left[(A_{k}u'_{k}+B_{k}u'^{*}_{k})(A^{*}_{k}u'^{*}_{k}+B^{*}_{k}u'_{k})
+k^{2}(A_{k}u_{k}+B_{k}u^{*}_{k})(A_{k}u_{k}+B_{k}u^{*}_{k})\right].
\end{equation}
For calculation of the above energy density, we consider the initial modes as follows
\begin{equation}
 \label{vac51}
u_{k}=\frac{e^{-ik\tau}}{\sqrt{k}}\left(1-\frac{i\alpha}{k\tau}-\frac{\beta}{k^{2}\tau^{2}}\right),
\end{equation}
and $\tau$ in term of Hankle index $\mu$ and slow roll parameter $\epsilon$ as \cite{asl55},
\begin{equation}
 \label{eta33} \tau=\frac{-1}{aH}\left(\frac{1}{1-\epsilon}\right)=\frac{-1}{aH}\left({\mu-\frac{1}{2}}\right).
\end{equation}
If we ignore the terms higher order than $1/(k\tau)^{2}$, we obtain \cite{man66},
$$\delta\rho_{ED}\sim\frac{1}{a^{4}(\tau)}\int\frac{d^{3}k}{(2\pi)^{3}}$$
\begin{equation}
 \label{vac60}
\times{k}\left[|A_{k}|^{2}+|B_{k}|^{2}+A_{k}B^{*}_{k}(\frac{2\alpha^{2}}{k^{2}\tau^{2}}
+\frac{2i\alpha}{k\tau}-1)e^{-2ik\tau}+B_{k}A^{*}_{k}{(\frac{2\alpha^{2}}{k^{2}\tau^{2}}-\frac{2i\alpha}{k\tau}-1)}e^{+2ik\tau}\right].
\end{equation}
If we consider the correction terms in (\ref{vac60}), these terms are of order $H/M_{Pl}$ and $(H/M_{Pl})^{2}$, that are very small for consideration \cite{man66}.\\
In the far past time limit $\tau\rightarrow{-\infty}$, we have dS mode with $\nu=3/2$ and we obtain from (\ref{vac60}) the following result
\begin{equation}
 \label{vac61}
\delta\rho_{ED}\sim\frac{1}{a^{4}(\tau)}\int\frac{d^{3}k}{(2\pi)^{3}}\left[|A_{k}|^{2}+|B_{k}|^{2}-A_{k}B^{*}_{k}e^{-2ik\tau}-B_{k}A^{*}_{k}e^{+2ik\tau}\right]k,
\end{equation}
this result complectly is similar to result \cite{tan1}. As is known well, this expression is divergent for any choice of $A_{k}$ and $B_{k}$. The finite vacuum energy density can be given by \cite{tan1, ash2}
\begin{equation}
 \label{vac61}
(\delta\rho_{ED})_{ren}\sim\frac{1}{a^{4}(\tau)}\int\frac{d^{3}k}{(2\pi)^{3}}{k}{|B_{k}|^{2}}.
\end{equation}
To make sure that the back-reaction is small, $B_{k}$ should be satisfied the following observational condition \cite{ash2},
\begin{equation}
\int\frac{d^{3}k}{(2\pi)^{3}}{k}{|B_{k}|^{2}}\ll{\epsilon\eta{H}^{2}M^{2}_{Pl}},
\end{equation}
where $\epsilon$ and $\eta$ are the slow-roll parameters.

\section{Calculation of Power Spectrum}
\subsection{Power Spectrum with Bunch-Davies Mode}
To calculate the power spectrum, we need to compute the following quantity \cite{per7, non29},
\begin{equation}
 \label{pow16} \langle\hat{u}_{k}(\tau)\hat{u}_{k'}(\tau)\rangle=\langle0|\hat{u}_{k}(\tau)\hat{u}_{k'}(\tau)|0\rangle=\frac{1}{2}(2\pi)^{3})\delta^{3}(k+k')|u_{k}(\tau)|^2,
\end{equation}
where we use (\ref{mod9}). Next, we should introduce some standard quantities in terms of curvature perturbation ${\cal R}_{k}(\tau)$,
\begin{equation}
 \label{pow17} \langle\hat{{\cal R}}_{k}(\tau)\hat{{\cal R}}_{k'}(\tau)\rangle=(2\pi)^{3}\delta^{3}(k+k')P_{{\cal R}},
\end{equation}
\begin{equation}
 \label{pow18} \Delta_{{\cal R}}^{2}=\frac{k^3}{2\pi^{2}}P_{{\cal R}},
\end{equation}
where
\begin{equation}
 \label{pow19} {\cal R}_{k}(\tau)=\frac{u_{k}(\tau)}{z}=\frac{u_{k}(\tau)}{a}(\frac{H}{{\dot{\bar{\phi}}}}).
\end{equation}
$P_{{\cal R}}$ is the power spectrum and $\Delta_{{\cal R}}^{2}$ is the dimensionless power spectrum \cite{non29}.\\
As we know, the scale dependence of the spectra follows from the \emph{time dependence of the Hubble parameter}\footnote{As it is argued in this paper, the excited-dS modes is applicable for the time-dependent Hubble parameter, so it is certainly expected scale dependence of the spectrum resulting from these nontrivial modes.} and is quantified by tilt \cite{per7},
\begin{equation}
 \label{spe20} n_{s}-1=\frac{d \ln{\Delta_{{\cal R}}^{2}}}{d \ln{k}}. \end{equation}
By using equations (\ref{pow17} -\ref{spe20}), in the super-horizon limit ($k\tau\ll{1}$), we have scale-invariant power spectrum for BD mode as,
\begin{equation}
 \label{inv2l} n_{s}=1 ,\quad       \Delta_{{\cal R}}^{2}=\frac{H^2}{(2\pi)^{2}}(\frac{H^2}{\dot{\bar{\phi}}^2}).
  \end{equation}
where $H$ is the Hubble constant for this scale-invariant power spectrum ( i.e. $n_{s}=1$) during exponential inflation in pure dS phase. But observations of CMB and LSS tell us conclusively that the power spectrum of the fluctuations produced during inflation is almost scale-invariant ( i.e. $n_{s}\approx1$) \cite{obs11, obs12}. Therefore, we believe that the geometry of the very early universe is almost dS (or excited-dS) with time-dependent Hubble parameter. In the next section, we replace nontrivial ED modes instead BD mode to obtain deviation from scale invariance in terms of $\nu$ and slow-roll parameter $\epsilon$.
\subsection{Power Spectrum with excited-de Sitter Modes}
For ED modes (\ref{mod28}), we have
\begin{equation}
 \label{pow31} P_{{\cal R}}=\frac{1}{2a^2}(\frac{H^2}{\dot{\bar{\phi}}^2})|u_{k}^{ED}(\tau)|^{2}.
\end{equation}
For super-horizon limit $k\tau\ll1$, we obtain modified power spectrum in the following general form:
$$ \Delta_{{\cal R}}^{2}=\frac{H^2}{(2\pi)^{2}}(\frac{H^2}{\dot{\bar{\phi}}^2})(1-\epsilon)^2\left[\alpha+\frac{\beta^2}{k^{2}\tau^{2}}\right]
$$
\begin{equation}
\label{del32} =\frac{H^2}{(2\pi)^{2}}(\frac{H^2}
 {\dot{\bar{\phi}}^2})(\frac{2}{2\nu-{1}})^2\left[\alpha+\frac{\beta^2}{k^{2}\tau^{2}}\right].
\end{equation}
For the last equation we use the (\ref{eta33}).\\
Here, to obtain the above relation, we assume the slow-roll parameter $ \epsilon $ is a constant \cite{asl55}. For special cases of $\nu$, we obtain as follows:
\begin{itemize}
  \item For the $\nu=\frac{1}{2}$, we have  $ \alpha=0$. So, the power spectrum for this case is vanish.
  \item In the pure dS phase or $\nu=\frac{3}{2}$, we obtain
    \begin{equation}
 \label{alf Bunch mode3/2} \alpha=1,\quad  \epsilon=\frac{-\dot{H}}{H^{2}}=0.
\end{equation}
So, we have BD mode and (\ref{del32}) reduced to the standard scale-invariant power spectrum (\ref{inv2l}) for this particular case.
  \item Also, if we consider $\nu=\frac{3}{2}+\frac{1}{p-1}$ \cite{asl55}, we have PLI with power $ p =\frac{1}{\epsilon}$ with $\epsilon<1$ and we consider for this case,
  \begin{equation}
 \label{alf Bunch mode3/2+} \alpha\neq1,\quad  \epsilon={Constant}.
\end{equation}
   So, for ED modes (\ref{mod28}) up to second order of $1/k\tau$, the modified power spectrum (\ref{del32}) change to the following result in terms of $\nu$,
  \begin{equation} \label{del36}
 \Delta_{{\cal R}}^{2}=\frac{H^2}{(2\pi)^{2}}(\frac{H^2}{\dot{\bar{\phi}}^2})\big[\frac{2\nu+1}{2(2\nu-1)}+(2\nu+1)^{2} \frac{(4\nu^{2}-9)^{2}}{64k^{2}\tau^{2}}\big],
\end{equation}
by considering the trans-Planckian effect that appears as a fixed scale $\tau=\tau_{0}$ \cite{tra49} and equations (\ref{del32}), the equation (\ref{del36}) change to,
\begin{equation}\label{del37}
 \Delta_{{\cal R}}^{2}=\frac{H^2}{(2\pi)^{2}}(\frac{H^2}{\dot{\bar{\phi}}^2})\left[\frac{2\nu+1}{2(2\nu-1)}+(\frac{2\nu+1}{2\nu-1})^{2} (\frac{4\nu^{2}-9}{4})^{2}(\frac{H}{M_{Pl}})^{2}\right].
\end{equation}
Note that for a given $k$ a finite $\tau_0$ is chosen in which the physical momentum corresponding to $k$ is given by some fixed scale $M_{Pl}$, where $ \tau_{0}=-\frac{M_{Pl}}{Hk} $ has a finite value. The trans-Planckian corrections in (\ref{del37})  are of order $(\frac{H}{M_{Pl}})^{2} $. In \cite{tra48} by using of effective field theory the similar correction has been obtained. Also, for the careful analysis of the different order of trans-Planckian corrections can be checked paper \cite{bra57}. \\
 Equations (\ref{del36}) and (\ref{del37}) indicate that whatever the value of $ \nu $ is closer to 3/2( or $ \epsilon\rightarrow0$), we have slow-roll PLI instead PLI and the deviation from scale-invariant spectrum becomes less and less. However in general, our excited-dS modes (\ref{mod28}) build a scale-dependent power spectrum during PLI. As well as because of obtained second order trans-Planckian correction in (\ref{del37}), it is possible that ED modes create non-Gaussian effects in CMB similar to effective field theory method \cite{Sen60}. We will investigate this issue in detail in the future works .\\
  \end{itemize}

\section{Conclusions}
Resent data from Planck and WMAP motivated us to deviation from BD mode and consider some nontrivial modes instead of BD mode. Based in this fact, we have studied the ED modes instead BD mode as the nontrivial initial states for primordial fluctuations. We have obtained ED modes by using asymptotic expansion of the Hankel functions in the general solution of the equation of motion for the field in the curved space-time.\\
Also, we have presented constraints of scalar spectral index of power spectrum and confirmed slow-roll PLI and ED modes are consistent with these observations. Then, we have showed the corrections in back-reaction associated with the non-linear terms of ED modes are very small for consideration. Finally, by making use of this nontrivial modes we have calculate the power spectrum for primordial scalar field fluctuations. Our result for power spectrum is scale-dependent and the corrections are up to the higher order of  $H/M_{Pl}$, consistent with the other recent works in this issue and for the dS space-time limit leaded to the standard scale-invariant power spectrum. In the future work, we will study the constraints on the general ED modes, primordial non-Gaussianity in the CMB and particle creation resulting from these non-trivial and non-linear initial states.

\acknowledgments

We would like to thank M.V. Takook, M. R. Tanhay and P. Pedram for useful and serious discussions. We thanks H. Firouzjahi, K. Nozari, S. Kundu and H. Pejhan for constructive conversations and comments. This work has been supported by the Islamic Azad University, Science and Research Branch, Tehran, Iran.


\end{document}